\documentclass[11pt]{article}
\usepackage{amsmath,amsfonts,amssymb}
\numberwithin{equation}{section}

\newcommand{\nc}{\newcommand}
\nc{\bib}{\bibitem}
\nc{\al}{\alpha}
\nc{\g}{\gamma}
\nc{\G}{\Gamma}
\nc{\D}{\Delta}
\nc{\eps}{\epsilon}
\nc{\la}{\lambda}
\nc{\La}{\Lambda}
\nc{\var}{\varphi}
\nc{\pa}{\partial}
\nc{\nn}{\nonumber \\ }
\nc{\be}{\begin{equation}}
\nc{\ee}{\end{equation}}
\nc{\bea}{\begin{eqnarray}}
\nc{\eea}{\end{eqnarray}}
\nc{\bra}[1]{\langle {#1}|}
\nc{\ket}[1]{|{#1}\rangle}
\nc{\gb}{\bar{g}}
\nc{\sbar}{\bar{s}}
\nc{\Ab}{\bar{A}}
\nc{\Db}{\bar{D}}
\nc{\Lc}{\mathcal{L}}
\nc{\Oc}{\mathcal{O}}
\nc{\Qh}{\hat{Q}}

\begin{document}

\topmargin -5mm
\oddsidemargin 5mm

\setcounter{page}{1}

\vspace{8mm}
\begin{center}
{\huge {\bf Isometry-preserving boundary conditions}}
\\[.3cm]
{\huge {\bf in the Kerr/CFT correspondence}}

\vspace{8mm}
 {\LARGE J{\o}rgen Rasmussen}
\\[.3cm]
 {\em Department of Mathematics and Statistics, University of Melbourne}\\
 {\em Parkville, Victoria 3010, Australia}
\\[.4cm]
 {\tt j.rasmussen@ms.unimelb.edu.au}

\end{center}

\vspace{8mm}
\centerline{{\bf{Abstract}}}
\vskip.4cm
\noindent
The near-horizon geometries of the extremal Kerr black hole and certain
generalizations thereof are considered.
Their isometry groups are all given by $SL(2,\mathbb{R})\times U(1)$.
The usual boundary conditions of the Kerr/CFT correspondence enhance the $U(1)$
isometry to a Virasoro algebra.
Various alternatives to these boundary conditions are explored.
Partial classifications are provided of the boundary conditions enhancing 
the $SL(2,\mathbb{R})$ isometries or separately the $U(1)$ isometry to a Virasoro algebra.
In the case of $SL(2,\mathbb{R})$-enhancing boundary conditions of a 
near-horizon geometry of the type considered, 
the conserved charges associated to the generators of the asymptotic Virasoro symmetry
form a centreless Virasoro algebra. 
\renewcommand{\thefootnote}{\arabic{footnote}}
\setcounter{footnote}{0}

\newpage

\section{Introduction}

Quantum gravity on three-dimensional anti-de Sitter ($AdS$) space was found in~\cite{BH86}
to be holographically dual to a two-dimensional conformal field theory (CFT).
In the spirit of this work, it was recently argued~\cite{GHSS0809} that the extremal
four-dimensional Kerr black hole~\cite{BH9905}, 
for which the angular momentum $J$ saturates the regularity bound
$J\leq GM^2$, is holographically dual to a chiral CFT in two dimensions.
This Kerr/CFT correspondence was subsequently~\cite{HMNS0811} 
generalized to a similar correspondence for the extremal Kerr-Newman black hole 
as well as for its $AdS$ and $dS$ generalizations. 
We refer to~\cite{MS9702} for earlier work on a dual description of the Kerr black hole, and 
to~\cite{wake1,wake1a,wake2,wake2a,wake3,wake3a,wake3b} 
for further progress in the wake of~\cite{GHSS0809}.

Excitations around the near-horizon extremal black holes can be controlled by imposing
appropriate boundary conditions. To every consistent set of boundary conditions,
there is an associated asymptotic symmetry group generated by the diffeomorphisms
obeying the conditions. The conserved charge of an asymptotic symmetry is constructed as a
surface integral and can be analyzed using the formalism of~\cite{BB0111,BC0708}
based on~\cite{BH86,AD82} and discussed extensively in~\cite{Com0708}.
An asymptotic symmetry transformation with vanishing conserved charge is rendered trivial.

The near-horizon metrics of the extremal black holes of our interest
all have an $SL(2,\mathbb{R})\times U(1)$ isometry group. 
In the studies~\cite{GHSS0809,HMNS0811} of the Kerr/CFT
correspondence and its generalizations, the $SL(2,\mathbb{R})$ becomes trivial
while the $U(1)$ is enhanced to a Virasoro algebra. This is in contrast to the situation in
studies of the G\"odel black hole~\cite{CD0701} and warped $AdS_3$~\cite{CD0808} 
in which an $SL(2,\mathbb{R})$ isometry is enhanced to a Virasoro algebra.

The boundary conditions imposed in~\cite{GHSS0809,HMNS0811}
are relevant for describing the ground-state entropy of the extremal
Kerr black hole or its generalization. A discussion of the microscopic origin of
the Bekenstein-Hawking entropy~\cite{Bek73} for a class of black holes
may be found in~\cite{SV9601} and references therein.

Once constructed, boundary conditions enhancing the $SL(2,\mathbb{R})$ isometry
of the extremal Kerr black hole, on the other hand, are  
speculated~\cite{GHSS0809} 
to be relevant for the understanding of the entropy of near-extremal fluctuations.
It is an objective of the present work to devise such 
boundary conditions and to study the resulting asymptotic symmetry group. 
The proposed boundary conditions are isometry-preserving in the sense that the original exact
$SL(2,\mathbb{R})$ isometries correspond to the global conformal transformations
(generated by the Virasoro modes $\ell_n$, $n=-1,0,1$) of the dual two-dimensional CFT.
Following the standard approach~\cite{BB0111,BC0708}, 
the conserved charges associated to these Virasoro-generating asymptotic Killing vectors 
are well-defined and non-vanishing, but yield a centreless Virasoro algebra. 
Challenges from so-called back-reaction effects are briefly indicated.

After reviewing the construction in~\cite{GHSS0809}, we argue that there
are infinitely many choices of boundary conditions yielding the same
centrally-extended Virasoro algebra as the $U(1)$-enhanced one
obtained in~\cite{GHSS0809}. We also show that there is a related class of boundary
conditions giving rise to a centreless $U(1)$-enhanced Virasoro algebra.
We then present a class of boundary conditions enhancing the $SL(2,\mathbb{R})$ isometries
to a centreless Virasoro algebra. 
The corresponding asymptotic symmetry is generated by an 
unusual differential-operator realization of the Virasoro algebra. 
A well-defined central extension of the algebra generated by the associated conserved charges 
is not permitted in this context -- not even when ignoring back-reaction effects. 
Nor does it seem possible, within our approach, to construct boundary
conditions resulting in two copies of the Virasoro algebra -- one enhancing the $U(1)$ isometry;
the other enhancing the $SL(2,\mathbb{R})$ isometries.
It certainly is possible, though, at the level of asymptotic Killing vectors, but the charges
associated to at least one of the two Virasoro copies are ill-defined or simply vanish. 
\\[.2cm]
While this work was being completed, the paper~\cite{MTY0907} on the Kerr/CFT
correspondence appeared. It shares some of our objectives 
and has an overlap in approach,
but is based on a particular choice of boundary conditions not considered here. 
As in our similar cases, the $SL(2,\mathbb{R})$ isometries of the extremal Kerr black hole 
are enhanced to a Virasoro algebra generated by a set of asymptotic Killing vectors.
Contrary to our cases, one verifies that the associated conserved charges all vanish, thus
rendering the corresponding asymptotic symmetries trivial.
The subsequent analysis of finite-temperature effects and 
entropy in~\cite{MTY0907} is based on {\em quasi-local} charges~\cite{BY9209}. 
Analyses of that kind are not carried out here.
A continuation of the work~\cite{MTY0907} can be found in~\cite{MTY0907b}.

\section{Near-horizon extremal geometry}

We are interested in the class~\cite{KLR0705} 
of extremal, stationary and rotationally symmetric four-dimensional
black holes whose near-horizon metric and gauge field are of the form
\bea
 d\sbar^2&=&\G(\theta)\Big(\!-r^2dt^2+\frac{dr^2}{r^2}+\al(\theta)d\theta^2\Big)
   +\g(\theta)\big(d\phi+krdt\big)^2\nn
 \Ab&=&f(\theta)\big(d\phi+krdt\big)
\label{ds2}
\eea
where $\theta\in[0,\pi]$ and $\phi\in[0,2\pi)$.
Among these, is the extremal Kerr-Newman black hole as well as its $AdS$ and
$dS$ generalizations. 
The corresponding isometry group $SL(2,\mathbb{R})\times U(1)$ is generated by
the Killing vectors
\be
 \big\{\pa_t,\ t\pa_t-r\pa_r,\ \big(t^2+\frac{1}{r^2}\big)\pa_t
   -2tr\pa_r-\frac{2k}{r}\pa_\phi\big\}\cup\big\{\pa_\phi\big\}
\label{iso}
\ee
In this work, we only consider the gravitational part but hope to
discuss the gauge transformations elsewhere. 

The near-horizon extremal Kerr metric, in particular, is obtained by setting
\be
 \G(\theta)=a^2(1+\cos^2\theta),\qquad 
 \g(\theta)=\frac{4a^2\sin^2\theta}{1+\cos^2\theta},\qquad
 \al(\theta)=k=1
\label{Kerr}
\ee
and the ADM mass and angular momentum of the extremal Kerr black hole are given by
\be
 M=\frac{a}{G},\qquad\quad J=\frac{a^2}{G}
\ee
The Kerr metric itself~\cite{Kerr63} describes a rotating black hole as a solution
to the four-dimensional vacuum Einstein equations.

\section{Asymptotic symmetry group}

We are interested in fluctuations of the near-horizon geometry of the extremal black hole
whose background metric $\gb_{\mu\nu}$ is defined in (\ref{ds2}).
We denote the corresponding perturbation of this metric by $h_{\mu\nu}$.
Asymptotic symmetries are generated by the diffeomorphisms whose action on the
metric generates metric fluctuations compatible with the chosen boundary conditions.
We are thus looking for contravariant vector fields $\eta$ along which the Lie derivative
of the metric is of the form 
\be
 \Lc_\eta \gb_{\mu\nu}\sim h_{\mu\nu}
\label{Lgh}
\ee
The asymptotic symmetry group is generated by the set of these transformations
modulo those whose charges, to be defined below, vanish.
The boundary conditions should therefore be strong enough to ensure well-defined
charges, yet weak enough to keep the charges non-zero. 

It is of interest to determine
the window of suitable boundary conditions. Once the asymptotic group
of a consistent set of boundary conditions has been determined, one may scan,
by weakening or strengthening the conditions, for other boundary conditions yielding the 
same or a closely related asymptotic group. Generally speaking, a strengthening of the 
boundary conditions will disallow certain asymptotic symmetries
but not allow new ones. A weakening of the boundary conditions, on the other hand,
will typically allow new asymptotic symmetries but may render some charges ill-defined.
A change in boundary conditions strengthening some parts of the metric fluctuations
but weakening others, may result in a different asymptotic symmetry group
or in an equivalent group obtained as an enhancement of different exact isometries.

To the asymptotic symmetry generator $\eta$ satisfying (\ref{Lgh}), 
one associates~\cite{BB0111,BC0708} the conserved charge
\bea
 Q_\eta&=&\frac{1}{8\pi G}\int_{\pa\Sigma}\sqrt{-\gb}k_\eta[h;\gb]
   =\frac{1}{8\pi G}\int_{\pa\Sigma}\frac{\sqrt{-\gb}}{4}\eps_{\al\beta\mu\nu}d_\eta^{\mu\nu}[h;\gb]
     dx^\al\wedge dx^\beta\nn
  &=&\frac{1}{16\pi G}\int_0^\pi\int_0^{2\pi}\sqrt{-\gb}
     \big(d_{\eta}^{tr}[h;\gb]-d_\eta^{rt}[h;\gb]\big)d\phi d\theta
\label{Q}
\eea
where
\be
 d_\eta^{\mu\nu}[h;\gb]
  =\eta^\nu \Db^\mu h-\eta^\nu \Db_\sigma h^{\mu\sigma} +\eta_\sigma\Db^\nu h^{\mu\sigma}
  -h^{\nu\sigma}\Db_\sigma\eta^\mu+\frac{1}{2}h\Db^\nu\eta^\mu
  +\frac{1}{2}h^{\sigma\nu}\big(\Db^\mu\eta_\sigma+\Db_\sigma\eta^\mu\big)
\label{dmunu}
\ee
and where $\pa\Sigma$ is the boundary of a three-dimensional
spatial volume, ultimately near spatial infinity. Here, indices are lowered and raised
using the background metric $\gb_{\mu\nu}$ and its inverse, 
$\Db_\mu$ denotes a background covariant derivative, 
while $h$ is defined as $h=\gb^{\mu\nu}h_{\mu\nu}$.
To be a well-defined charge in the asymptotic limit, 
the underlying integral must be finite as $r\to\infty$.
If the charge vanishes, the asymptotic symmetry is rendered trivial.
The algebra generated by the set of well-defined charges is governed
by the Dirac brackets computed~\cite{BB0111,BC0708} as
\be
 \big\{Q_\eta,Q_{\hat\eta}\big\}
  =Q_{[\eta,\hat\eta]}+\frac{1}{8\pi G}\int_{\pa\Sigma}\sqrt{-\gb}k_{\eta}[\Lc_{\hat\eta}\gb;\gb]
\label{QQ}
\ee
where the integral yields the eventual central extension.

\section{Kerr/CFT correspondence}

For ease of comparison, we here use the global coordinates of the near-horizon
extremal Kerr black hole used in~\cite{GHSS0809} in which the metric reads
\be
 d\sbar^2=2GJ\Omega^2\Big(\!-(1+r^2)d\tau^2+\frac{dr^2}{1+r^2}+d\theta^2
  +\Lambda^2\big(d\var+rd\tau\big)^2\Big)
\label{ds2Kerr}
\ee
where
\be
 \Omega^2=\Omega^2(\theta)=\frac{1+\cos^2\theta}{2},\qquad\qquad
  \La=\Lambda(\theta)=\frac{2\sin\theta}{1+\cos^2\theta}
\ee
In these coordinates, the rotational $U(1)$ isometry is generated by the Killing vector
$\pa_\var$, while the $SL(2,\mathbb{R})$ isometries do not concern us here.
Written in the ordered basis $\{\tau,r,\var,\theta\}$, the imposed boundary conditions read
\be
 h_{\mu\nu}=\Oc\!\left(\!\!\begin{array}{cccc} r^2&r^{-2}&1&r^{-1} \\ &r^{-3}&r^{-1}&r^{-2} \\ 
   &&1&r^{-1} \\ &&&r^{-1} \end{array} \!\!\right),
  \qquad\quad h_{\mu\nu}=h_{\nu\mu}
\label{hKerr}
\ee
To preserve extremality of the Kerr black hole, one imposes the additional
condition $Q_{\pa_\tau}=0$. The asymptotic Killing vectors are given by
\be
 K_\eps =\Oc(r^{-3})\pa_\tau
   +\big(\!-r\eps'(\var)+\Oc(1)\big)\pa_r
   +\big(\eps(\var)+\Oc(r^{-2})\big)\pa_{\var}
   +\Oc(r^{-1})\pa_{\theta}
\ee
where $\eps(\var)$ is a smooth function, in addition to the `trivialized' $\pa_\tau$. 
The generators of the corresponding asymptotic symmetry read
\be
 \xi=-r\eps'(\var)\pa_r+\eps(\var)\pa_{\var}
\ee 
and form the centreless Virasoro algebra 
\be
 \big[\xi_{\eps},\xi_{\hat\eps}\big]=\xi_{\eps\hat\eps'-\eps'\hat\eps}
\label{Virdiff}
\ee
The usual form of the Virasoro algebra 
is obtained by choosing an appropriate basis for the functions $\eps(\var)$ and $\hat\eps(\var)$.
This symmetry is an enhancement 
of the exact $U(1)$ isometry generated by the Killing vector $\pa_\var$ of 
(\ref{ds2Kerr}) as the latter is recovered by setting $\eps(\var)=1$. 
The associated charges are computed using
\be
 \sqrt{-\gb}k_\xi[h;\gb]=\left(\frac{1}{2}\eps'\La rh_{r\var}
   -\frac{1}{4}\eps\La\Big(\La^2\frac{h_{\tau\tau}}{r^2}
     +2r\pa_\var h_{r\var}+\big(\La^2+1\big)h_{\var\var}\Big)\right)d\var\wedge d\theta+\ldots
\label{kxiKerr}
\ee
where we have introduced the shorthand notation $\eps=\eps(\var)$.
The dots indicate that terms not contributing to the charge (\ref{Q}) have been omitted.
With respect to the basis $\xi_n(\var)$, where $\eps_n(\var)=-e^{-in\var}$, 
one introduces the dimensionless quantum versions of the conserved charges
\be
 L_n=\frac{1}{\hbar}\Big(Q_{\xi_n}+\frac{3J}{2}\delta_{n,0}\Big)
\label{L}
\ee
After the usual substitution $\{.,.\}\to-\frac{i}{\hbar}[.,.]$ of Dirac brackets by quantum commutators,
the quantum charge algebra is recognized as the 
centrally-extended Virasoro algebra~\cite{GHSS0809}
\be
 \big[L_n,L_m\big]=(n-m)L_{n+m}+\frac{c}{12}n(n^2-1)\delta_{n+m,0},\qquad\quad
   c=\frac{12J}{\hbar}
\label{VirKerr}
\ee

\subsection{Partial classification of $U(1)$-enhancing boundary conditions}

The Lie derivative along $\xi$ of the background metric $\gb_{\mu\nu}$ is given by
\be
 \Lc_\xi\gb_{\mu\nu}=-2GJ\Omega^2\left(\!\!\begin{array}{cccc} 
    2(\La^2-1)r^2\eps'&0&0&0 \\ 
   0&\frac{2}{(1+r^2)^2}\eps'&\frac{r}{1+r^2}\eps''&0 \\ 
   0&\frac{r}{1+r^2}\eps''&-2\La^2\eps'&0 
   \\ 0&0&0&0 \end{array} \!\!\right)
\label{Liexigb}
\ee
and has only four (independent) non-trivial entries. 
It is thus natural to ask if the boundary conditions
(\ref{hKerr}) can be strengthened while maintaining the Virasoro algebra (with its
non-trivial central charge (\ref{VirKerr})) 
as an enhancement of the $U(1)$ isometry generated by the Killing vector $\pa_\var$.  
The {\em strongest} such boundary conditions are
\be
 h_{\mu\nu}=\left(\!\!\begin{array}{cccc} \Oc(r^2)&0&0&0 \\ 0&\Oc(r^{-4})&\Oc(r^{-1})&0 \\ 
   0&\Oc(r^{-1})&\Oc(1)&0 \\ 0&0&0&0 \end{array} \!\!\right)
\label{hKerrStrongC}
\ee
and have asymptotic symmetries generated by $\xi$. The expression (\ref{kxiKerr})
remains and the Virasoro algebra generated by the associated charges $Q_\xi$ has
the same central charge as above. The additional condition $Q_{\pa_\tau}=0$ is still imposed.

Determining the {\em weakest} boundary conditions yielding a $U(1)$-enhanced
Virasoro algebra is more delicate. As already indicated, new symmetries may arise.
Even if the charges of the Virasoro algebra in question remain well-defined, the new
symmetries may be ill-defined and thus render the boundary conditions inconsistent. 
It is, a priori, unclear if such ill-defined charges can be dealt with by imposing additional
boundary conditions setting them equal to zero, as in the case $Q_{\pa_\tau}=0$ above.
It is beyond the scope of the present work to classify such possibilities.
Instead, we merely point out that the weakest boundary conditions 
keeping the Virasoro-generating charges {\em themselves} well-defined are of the form
\be
 h_{\mu\nu}=\Oc\!\left(\!\!\begin{array}{cccc} r^2&r^{p_{\tau r}}&r&r \\ 
   &r^{-2}&r^{-1}&r^{p_{r \theta}} \\ 
   &&1&1 \\ &&&1 \end{array} \!\!\right),
  \qquad\quad h_{\mu\nu}=h_{\nu\mu}
\label{hwea}
\ee
Here, $p_{\tau r}$ and $p_{r \theta}$ are real parameters bounded from above by the
applicability, at infinity, of the linear theory assumed in the charge formula 
(\ref{Q})~\cite{BC0708}. Explicit values for the bounds are not discussed here, though.
The general form of (\ref{hwea}) follows from a simple inspection of the $r$-powers in 
\bea
 \sqrt{-\gb}k_\xi[h;\gb]\big|_{d\var\wedge d\theta}&=&
  -\frac{\La^3\eps}{4(r^2+1)}h_{\tau\tau}+\frac{r\eps''+2r\La^4\eps-2(r^2+1)\La^2\eps\pa_r
    -2r\eps'\pa_\var}{4\La(r^2+1)}h_{\tau\var}\nn
  &-&\frac{r\eps'}{2(r^2+1)}\pa_\theta\big(\La h_{\tau\theta}\big)
   +\frac{\La^3}{4}(r^2+1)\eps h_{rr}
   +\frac{\La}{2}\big(r\eps'-r\eps\pa_\var+\eps\pa_t\big)h_{r\var}\nn
  &-&\frac{\La^2\big((\La^2+1)r^2+1\big)\eps-2\La^2r(r^2+1)\eps\pa_r-2r\eps'\pa_t}{4\La(r^2+1)}
    h_{\var\var}+\frac{r^2\eps'}{2(r^2+1)}\pa_{\theta}\big(\La h_{\var\theta}\big)\nn
  &-&\frac{\La}{4(r^2+1)}\big(\La^2(r^2+1)\eps-r^2\eps''+2r^2\eps'\pa_\var
    -2r\eps'\pa_t\big)h_{\theta\theta}
\eea 
where subleading terms have been included to show that $p_{\tau r}$ and $p_{r \theta}$
are unaffected by this particular evaluation. Compared with (\ref{Liexigb}), we see that  
the contributions from $\Lc_\xi\gb_{\tau\tau}$, $\Lc_\xi\gb_{r\var}$ and 
$\Lc_\xi\gb_{\var\var}$ to the central
charge are independent of the particular choice of boundary conditions considered here,
while $\Lc_\xi\gb_{rr}$ does not contribute to any of them.
The central charge (\ref{VirKerr}) is thus the same for all these choices.

\subsubsection{Centreless Virasoro algebra}

As we are about to demonstrate, 
one obtains a {\em centreless} Virasoro algebra as an enhancement
of the $U(1)$ isometry by imposing boundary conditions $h_{\mu\nu}$ in one of 
the three `ranges'
\be
 \left(\!\!\begin{array}{cccc} 0&0&\Oc(r)&0 \\ 
   0&0&0&0 \\ 
   \Oc(r)&0&\Oc(1)&0 \\ 0&0&0&0 \end{array} \!\!\right)
 \ <\ 
 \left(\!\!\begin{array}{cccc} h_{\tau\tau}&\Oc(r^{p_{\tau r}})&\Oc(r)&\Oc(r^{p_{\tau\theta}}) \\ 
   &h_{rr}&h_{r\var}&\Oc(r^{p_{r\theta}}) \\ 
   &&\Oc(1)&\Oc(r^{p_{\var\theta}}) \\ 
   &&&\Oc(1) \end{array} \!\!\right),\qquad\quad h_{\mu\nu}=h_{\nu\mu}
\ee
where 
\bea
 &&h_{\tau\tau},\ h_{rr},\ h_{r\var}=\Oc(r),\ \Oc(r^{-2}),\ \Oc(r^{-1})\nn
 \mathrm{or}\ &&  h_{\tau\tau},\ h_{rr},\ h_{r\var}=\Oc(r^2),\ \Oc(r^{-5}),\ \Oc(r^{-1})\nn
 \mathrm{or}\ &&  h_{\tau\tau},\ h_{rr},\ h_{r\var}=\Oc(r^2),\ \Oc(r^{-2}),\ \Oc(r^{-2})
\eea
in addition to $Q_{\pa_\tau}=0$. As in the discussion following (\ref{hwea}), the real parameters
$p_{\tau r}$, $p_{\tau\theta}$, $p_{r\theta}$ and $p_{\var\theta}$ are bounded from above. 
Either of the new conditions $h_{\tau\tau}=\Oc(r)$,
$h_{r\var}=\Oc(r^{-2})$ or $h_{rr}=\Oc(r^{-5})$ reduces the symmetry generator $\xi$ from 
$\xi=-r\eps'(\var)\pa_r+\eps(\var)\pa_{\var}$ (as allowed by boundary conditions in the 
range from (\ref{hKerrStrongC}) to (\ref{hwea}))
to $\xi=\eps(\var)\pa_{\var}$. Further conditions may have to be imposed to ensure
finiteness of eventual charges different from the ones
generated by the new asymptotic symmetry $\xi=\eps(\var)\pa_\var$, but this question is not
addressed here. Along $\xi=\eps(\var)\pa_\var$, 
the Lie derivative of the background metric reads
\be
 \Lc_{\xi}\gb_{\mu\nu}=2GJ\Omega^2\La^2\eps'\left(\!\!\begin{array}{cccc} 
    0&0&r&0 \\ 
   0&0&0&0 \\ 
   r&0&2&0 
   \\ 0&0&0&0 \end{array} \!\!\right)
\ee
The associated conserved and central charges follow from
\bea
 \!\!\!\!\!\!\sqrt{-\gb}k_\xi[h;\gb]\big|_{d\var\wedge d\theta}&=&
  -\frac{\La^3\eps}{4(r^2+1)}h_{\tau\tau}
  +\frac{\La\eps}{2}\Big(\frac{\La^2r}{r^2+1}-\pa_r\Big)h_{\tau\var}
  +\frac{\La^3}{4}(r^2+1)\eps h_{rr}\nn
 &+&\frac{\La}{4}\big(r\eps'-2r\eps\pa_\var+2\eps\pa_t\big)h_{r\var}
  -\frac{\La\eps}{4}\Big(\frac{\La^2r^2}{r^2+1}+1-2r\pa_r\Big)h_{\var\var}
  -\frac{\La^3}{4}\eps h_{\theta\theta}
\eea 
and
\be
 \frac{1}{8\pi G}\int_{\pa\Sigma}\sqrt{-\gb}k_{\xi}[\Lc_{\hat\xi}\gb;\gb]
   =-\frac{J}{\pi}\int_0^{2\pi}\eps(\var)\hat\eps'(\var)d\var
\ee
where $\xi=\eps(\var)\pa_\var$ and $\hat\xi=\hat\eps(\var)\pa_\var$.
Using the same basis $\xi_n(\var)$ as above, but with
\be
 L_n=\frac{1}{\hbar}\big(Q_{\xi_n}+J\delta_{n,0}\big)
\label{L2}
\ee
the quantum charge algebra is recognized as the centreless Virasoro algebra
\be
 \big[L_n,L_m\big]=(n-m)L_{n+m}
\ee

\section{Enhancing the $SL(2,\mathbb{R})$ isometries}

Returning to the general near-horizon geometry of an extremal black hole
whose background metric $\gb_{\mu\nu}$ is defined in (\ref{ds2}),
we are now looking for fluctuations $h_{\mu\nu}$ of the background metric enhancing the
$SL(2,\mathbb{R})$ isometries to a Virasoro algebra. That is, some of the generators of 
the asymptotic symmetry group should correspond to the generators of the $SL(2,\mathbb{R})$ 
isometries. Also, since these fluctuations are expected to be relevant for the description
of {\em near-extremal} perturbations, we refrain from imposing $Q_{\pa_t}=0$.
Aside from this {\em weakening} of the boundary conditions, it is natural to
expect otherwise {\em stronger} boundary conditions than the ones enhancing
the $U(1)$ isometry.
To select such boundary conditions, we reverse-engineer the problem.

First, though, we note that supplementing the boundary conditions (\ref{hKerr})
with the condition $Q_{\pa_t}=0$ corresponds to restricting to solutions with vanishing energy.
This zero-energy condition was imposed in~\cite{GHSS0809} not only to preserve 
extremality and study the ground states of the Kerr black hole, 
but also to ensure finiteness of the associated conserved charges.
It was subsequently argued in~\cite{wake3a} that this additional condition actually follows from
the boundary conditions (\ref{hKerr}).
It was also argued that so-called ``back-reaction effects" at orders higher than linear
could impose vanishing conditions, known as ``linearization-stability constraints"\!, 
on seemingly well-defined charges.
In particular, boundary conditions admitting near-extremal perturbations of the
near-horizon extremal Kerr geometry preserving any of the $SL(2,\mathbb{R})$ isometries
are believed to back-react so strongly that the Kerr asymptotics would break down.
Despite these assertions, we find it worthwhile to continue our `linear' analysis of 
$SL(2,\mathbb{R})$-enhancing boundary conditions of the near-horizon geometry (\ref{ds2}). 

Thus, we now consider one of the simplest possible sets of asymptotic Killing vectors
generating a Virasoro algebra whose `global' subalgebra 
(generated by $\ell_n$, $n=-1,0,1$) corresponds to the $SL(2,\mathbb{R})$ isometries, namely
\be
 K_\eps=\big[\eps(t)+\frac{\eps''(t)}{2r^2}+\Oc(r^{-4})\big]\pa_t
  +\big[-r\eps'(t)+\Oc(r^{-1})\big]\pa_r  
  +\big[-\frac{k\eps''(t)}{r}+\Oc(r^{-3})\big]\pa_{\phi}
  +\big[\Oc(r^{-2})\big]\pa_\theta
\label{xidiff}
\ee
Here, $\eps(t)$ is a smooth function and it follows that
\be
 \big[K_\eps,K_{\hat\eps}\big]=K_{\eps\hat\eps'-\eps'\hat\eps}
\ee
The associated asymptotic symmetry generator is given by the contravariant vector field
\be
 \kappa_\eps=\big(\eps(t)+\frac{\eps''(t)}{2r^2}\big)\pa_t
  -r\eps'(t)\pa_r-\frac{k\eps''(t)}{r}\pa_{\phi}
\label{kappa}
\ee
and the three $SL(2,\mathbb{R})$ 
generators in (\ref{iso}) follow by setting $\eps(t)=t^{n+1}$, $n=-1,0,1$. 
The Lie derivative along $\kappa_\eps$ of the background metric is given by
\be
 \Lc_{\kappa_\eps}\gb_{\mu\nu}=-\eps'''\left(\!\!\begin{array}{cccc} 
    \G+k^2\g&0&\frac{k\g}{2r}&0 \\ 
   0&0&0&0 \\ 
   \frac{k\g}{2r}&0&0&0 
   \\ 0&0&0&0 \end{array} \!\!\right)
\ee
here written in the ordered basis $\{t,r,\phi,\theta\}$ and in terms of the shorthands
\be
 \G=\G(\theta),\qquad\g=\g(\theta),\qquad\al=\al(\theta)
\ee
A set of boundary conditions compatible with these diffeomorphisms are
\be
 h_{\mu\nu}=\Oc\!\left(\!\!\begin{array}{cccc} 1&r^{-3}&r^{-1}&r^{-2} \\ 
   &r^{-4}&r^{-3}&r^{-3} \\ 
   &&r^{-2}&r^{-2} \\ &&&r^{-2} \end{array} \!\!\right),
  \qquad\quad h_{\mu\nu}=h_{\nu\mu}
\label{hnaive}
\ee

There are several problems with this construction. 
First, the associated conserved charges $Q_{\kappa_{\eps}}$ vanish, thereby rendering 
the asymptotic symmetries trivial.  
Second, the asymptotic symmetry generators (\ref{kappa}) do not quite form a Virasoro
algebra as we have
\be
 [\kappa_\eps,\kappa_{\hat\eps}]=\kappa_{\eps\hat\eps'-\eps'\hat\eps}
  +\frac{\eps''(t)\hat\eps'''(t)-\eps'''(t)\hat\eps''(t)}{4r^4}\big(\pa_t-2kr\pa_\phi\big)
\label{comm}
\ee
A proper differential-operator realization of the Virasoro algebra containing the
$SL(2,\mathbb{R})$ isometries (\ref{iso}) {\em does} exist, though,
and is discussed in the following. The issue with the charges
is subsequently addressed and resolved, and we are ultimately left with a
well-defined and non-vanishing set of conserved charges realizing the Virasoro algebra.
As we will see, the symmetry generators $\kappa_\eps$ (\ref{kappa}) differ from
the proper Virasoro generators by diffeomorphisms rendered trivial
by their vanishing charges.

\subsection{Asymptotic Virasoro symmetry}
\label{SecVir}

For every smooth function $\eps(t)$, we introduce the contravariant vector field
\be
 \xi=\xi^\mu\pa_\mu=\cosh\cdot\eps(t)\pa_t-r^2\sinh\cdot\eps(t)\pa_r-k\sinh\cdot\eps'(t)\pa_\phi
\label{xi}
\ee 
where
\bea
 &&\cosh\cdot\eps(t):=\cosh\big(\frac{\pa_t}{r}\big)\eps(t)
  =\sum_{n=0}^{\infty}\frac{1}{(2n)!r^{2n}}\eps^{(2n)}(t)\nn
 &&\sinh\cdot\eps(t):=\sinh\big(\frac{\pa_t}{r}\big)\eps(t)
  =\sum_{n=0}^{\infty}\frac{1}{(2n+1)!r^{2n+1}}\eps^{(2n+1)}(t)
\eea
with the $n$'th derivative of $\eps(t)$ denoted by $\eps^{(n)}(t)$.
We note that
\be
 \pa_t\xi^r=r^4\pa_r\xi^t,\qquad\quad
 \pa_t\xi^t=r^2\pa_r\big(\frac{\xi^r}{r^2}\big),\qquad\quad
 \xi^\phi=\frac{k}{r^2}\pa_t\xi^r
\ee
After a bit of algebra, one verifies that the vectors (\ref{xi}) satisfy (\ref{Virdiff}), thus providing a 
somewhat unusual differential-operator realization of the centreless Virasoro algebra.
With $\xi$ as the candidate for the generator of the corresponding asymptotic symmetry,
we now continue the reverse-engineered selection of suitable
$SL(2,\mathbb{R})$-enhancing boundary conditions.

Along $\xi$, the Lie derivative of the background metric is worked out to be
\be
 \Lc_{\xi}\gb_{\mu\nu}=\!\left(\!\!\begin{array}{cccc} 
    2r\big(k\g \pa_t\xi^\phi-(\G-k^2\g)(\xi^r+r\pa_t\xi^t)\big)&
    k\g r\big(kr\pa_r\xi^t+\pa_r\xi^\phi\big)&
    \g\big(k(\xi^r+r\pa_t\xi^t)+\pa_t\xi^\phi\big)&0 \\ 
   k\g r\big(kr\pa_r\xi^t+\pa_r\xi^\phi\big)&
   \frac{2\G}{r}\pa_r\big(\frac{\xi^r}{r}\big)&
   \g\big(kr\pa_r\xi^t+\pa_r\xi^\phi\big)&0 \\ 
   \g\big(k(\xi^r+r\pa_t\xi^t)+\pa_t\xi^\phi\big)&
   \g\big(kr\pa_r\xi^t+\pa_r\xi^\phi\big)&0&0 
   \\ 0&0&0&0 \end{array} \!\!\right)
\ee
Its non-vanishing components can be written as
\bea
 \Lc_\xi\gb_{tt}&=&-4\sum_{n=0}^{\infty}r^{-2n}\frac{n+1}{(2n+3)!}\big(\G+2(n+1)k^2\g\big)
    \eps^{(2n+3)}(t)\nn
 \Lc_\xi\gb_{tr}=\Lc_\xi\gb_{rt}
   &=&2k^2\g\sum_{n=0}^{\infty}r^{-3-2n}\frac{n+1}{(2n+3)!}\eps^{(2n+4)}(t)\nn
 \Lc_\xi\gb_{t\phi}=\Lc_\xi\gb_{\phi t}
   &=&-4k\g\sum_{n=0}^{\infty}r^{-1-2n}\frac{(n+1)^2}{(2n+3)!}\eps^{(2n+3)}(t)\nn
 \Lc_\xi\gb_{rr}&=&4\G\sum_{n=0}^{\infty}r^{-4-2n}\frac{n+1}{(2n+3)!}\eps^{(2n+3)}(t)\nn
 \Lc_\xi\gb_{r\phi}=\Lc_\xi\gb_{\phi r}
   &=&2k\g\sum_{n=0}^{\infty}r^{-4-2n}\frac{n+1}{(2n+3)!}\eps^{(2n+4)}(t)  
\eea
and it is observed that
\be
 \Lc_\xi\gb_{tt}+r^4\Lc_\xi\gb_{rr}=2kr\Lc_\xi\gb_{t\phi},\qquad\quad
 \Lc_\xi\gb_{tr}=kr\Lc_\xi\gb_{r\phi}
\label{Lgid}
\ee
It is straightforward, albeit rather tedious, to compute the relevant part of the integrand
in the surface integral (\ref{Q}) defining $Q_\xi$, and we find
\bea
 \sqrt{-\gb}k_\xi[h;\gb]\big|_{d\phi\wedge d\theta}&=&\sqrt{\frac{\al}{4\g\G}}\Big(
  \frac{k^2\g^2}{2\G}\pa_r(r\xi^t)\big(r^3h_{rr}-\frac{h_{tt}}{r}\big)
   +H_{t\phi}+H_{r\phi}+H_{\phi\phi}+H_{\theta\theta}\Big)\nn
 &+&\frac{1}{2r^2}\pa_\theta\Big(\sqrt{\frac{\g}{\al\G}}\big(r^4\xi^t h_{r\theta}
    +\xi^r(h_{t\theta}-krh_{\phi\theta})\big)\Big)
  +\pa_\phi \Phi
\label{2h}
\eea
where
\bea
 H_{t\phi}&=&\Big(\frac{k^3\g^2}{\G}\pa_r(r\xi^t)+\frac{\g}{2r}\pa_r(r\xi^\phi)\Big)h_{t\phi}
   -k\g r\pa_r(r\xi^t)\pa_r h_{t\phi}\nn
 H_{r\phi}&=&-\frac{\g}{2}\Big(kr^2\pa_r\big(\frac{\xi^r}{r}\big)+\pa_t\xi^\phi\Big)h_{r\phi}
   +k\g r\pa_r(r\xi^t)\pa_t h_{r\phi}\nn
 H_{\phi\phi}&=&\Big(\big(1+\frac{k^2\g}{2\G}\big)(\G-k^2\g)r\pa_r(r\xi^t)
    -\frac{k\g}{2}\pa_r(r\xi^\phi)\Big)h_{\phi\phi}-\frac{\G\xi^r}{r^2}\pa_t h_{\phi\phi}\nn
   &-&\big(\G\xi^t-k^2\g\pa_r(r\xi^t)\big)r^2\pa_r h_{\phi\phi}\nn
 H_{\theta\theta}&=&\frac{\g}{\al}\Big(\big(1-\frac{k^2\g}{2\G}\big)r\pa_r(r\xi^t)h_{\theta\theta}
   -\frac{\xi^r}{r^2}\pa_t h_{\theta\theta}-r^2\xi^t\pa_r h_{\theta\theta}\Big)
\eea
and
\be
 \Phi=\sqrt{\frac{\al}{4\g\G}}\Big(\frac{\G}{r^2}\xi^r h_{t\phi}
   +\big(\G\xi^t-k^2\g\pa_r(r\xi^t)\big)r^2h_{r\phi}+\frac{k\g\xi^r}{\al r}h_{\theta\theta}\Big)
\ee
It is emphasized that these expressions are valid for all $r$, and we note that (\ref{2h})
is independent of $h_{tr}=h_{rt}$. 
The total $\phi$-derivative $\pa_\phi\Phi$ can be ignored in the surface integral (\ref{Q}).

\subsection{Boundary conditions}

We initially require the boundary conditions to be of the form
\be
 h_{\mu\nu}=\Oc(r^{p_{\mu\nu}}),  \qquad\quad h_{\mu\nu}=h_{\nu\mu}
\label{hO}
\ee
where we allow $p_{\mu\nu}=-\infty$ for some coordinates, as for several entries
in (\ref{hKerrStrongC}), for example. 
The {\em strongest} such conditions compatible with the Virasoro symmetry generator
$\xi$ (\ref{xi}) are given by
\be
 h_{\mu\nu}=\Oc\!\left(\!\!\begin{array}{cccc} 1&r^{-3}&r^{-1}&0 \\ 
   &r^{-4}&r^{-4}&0 \\ 
   &&0&0 \\ &&&0 \end{array} \!\!\right),
  \qquad\quad h_{\mu\nu}=h_{\nu\mu}
\label{hs}
\ee
but one verifies that the associated conserved charges vanish.

The {\em weakest} boundary conditions (\ref{hO}), compatible with $\xi$ and
yielding well-defined charges $Q_\xi$, are given by
\be
 h_{\mu\nu}=\Oc\!\left(\!\!\begin{array}{cccc} r&r^{p_{tr}}&1&r \\ 
   &r^{-3}&r^{-1}&r^{-2} \\ 
   &&r^{-1}&1 \\ &&&r^{-1} \end{array} \!\!\right),
   \qquad\quad p_{tr}\geq-3,
  \qquad\quad h_{\mu\nu}=h_{\nu\mu}
\label{hw}
\ee
These bounds on the asymptotic boundary conditions follow from
analyzing the leading terms in (\ref{2h}) given by
\bea
 \sqrt{-\gb}k_\xi[h;\gb]\big|_{d\phi\wedge d\theta}&=&\eps\sqrt{\frac{\al}{4\g\G}}\Big(
  \frac{k^2\g^2}{2\G}\big(r^3 h_{rr}-\frac{h_{tt}}{r}\big)+\frac{k^3\g^2}{\G}h_{t\phi}
    +k\g r(\pa_t h_{r\phi}-\pa_r h_{t\phi})\nn
  &&+(\G-k^2\g)\big((1+\frac{k^2\g}{2\G})rh_{\phi\phi}
    -r^2\pa_r h_{\phi\phi}\big)
    +\frac{\g}{\al}\big(1-\frac{k^2\g}{2\G}\big)rh_{\theta\theta}
    -\frac{\g r^2}{\al}\pa_r h_{\theta\theta}\Big)\nn
  &+&\frac{1}{2}\pa_\theta\Big(\sqrt{\frac{\g}{\al\G}}\big(\eps r^2 h_{r\theta}
    -\eps'\big(\frac{h_{t\theta}}{r}-kh_{\phi\theta}\big)\big)\Big)
 +\ldots
\eea
where the total $\phi$-derivatives have been ignored. Including these derivatives 
might prompt one to strengthen the allowed fluctuation $h_{r\phi}$ 
unnecessarily from $\Oc(r^{-1})$ to $\Oc(r^{-2})$. 
As in the discussion following (\ref{hwea}), the real parameter $p_{tr}$ is bounded from
above by the linear theory underlying (\ref{Q}).

The conserved charges $Q_\xi$ corresponding to boundary conditions in the range
from (\ref{hs}) to (\ref{hw}) are non-zero if 
at least one of the bounds in (\ref{hw}), $h_{tr}$ excluded, is saturated. 
In all such cases, the charges generate a {\em centreless} Virasoro algebra
since a simple $r$-power counting asymptotically gives 
$\Lc_\xi\gb_{\mu\nu}<h_{\mu\nu}$ for all $\mu,\nu$.

With reference to the comments following (\ref{comm}), we note that 
$Q_{\xi_\eps-\kappa_\eps}=0$ for all boundary conditions in the range from
(\ref{hs}) to (\ref{hw}). This implies the announced triviality of the difference
between the naive symmetry generators $\kappa_\eps$ and the proper Virasoro
generators $\xi_\eps$ for every smooth function $\eps(t)$.

Many alternatives exist to boundary conditions of the simple type (\ref{hO}). 
Imposing the separate condition $h_{t\theta}=krh_{\phi\theta}$, for example,
renders (\ref{2h}) independent of $h_{t\theta}$ and $h_{\phi\theta}$
thus weakening the conditions on the real parameter $p_{\phi\theta}$ in 
$h_{\phi\theta}=\Oc(r^{p_{\phi\theta}})$, and subsequently in $h_{t\theta}=\Oc(r^{p_{t\theta}+1})$.
Imposing conditions resembling the relations (\ref{Lgid}) is another interesting possibility.

We emphasize that there are infinitely many sets of consistent boundary
conditions simultaneously admitting two copies of Virasoro-generating 
asymptotic Killing vectors enhancing the $SL(2,\mathbb{R})$ and $U(1)$ 
isometries separately. Within the realm of boundary conditions considered here,
however, at least one of these copies gives rise to vanishing or ill-defined charges,
and we are left with {\em at most} one quantum charge Virasoro algebra.

\subsection{Spurious asymptotic symmetries}

There may be many more asymptotic Killing vectors and asymptotic symmetry generators
than conserved charges since the surface integrals (\ref{Q}) 
producing the charges from the generators may vanish. The corresponding 
contravariant vector fields thus generate {\em spurious} asymptotic symmetries.
Let us illustrate this by considering the diffeomorphisms whose action on the background
metric generate fluctuations compatible with
\be
  h_{\mu\nu}=\left(\!\!\begin{array}{cccc} \Oc(r^{-2\ell_0})&0&\Oc(r^{-1-2\ell_0})&0 \\ 
    0&\Oc(r^{-4-2\ell_0})&0&0 \\ 
   \Oc(r^{-1-2\ell_0})&0&0&0 \\ 0&0&0&0 \end{array} \!\!\right)
\label{hrr}
\ee
for $\ell_0$ a non-negative integer. Since these conditions are stronger than (\ref{hs}),
the diffeomorphisms to be discussed are compatible with all boundary conditions
in the range from (\ref{hs}) to (\ref{hw}).
For every integer $\ell\geq\ell_0+2$, we find the contravariant vector field
\be
 \eta_{(2\ell)}=\frac{\rho_{(2\ell)}'(t)}{2\ell r^{2\ell}}\pa_t
   -\frac{\rho_{(2\ell)}(t)}{r^{2\ell-3}}\pa_r
   -\frac{k\rho_{(2\ell)}'(t)}{(2\ell-1)r^{2\ell-1}}\pa_\phi
\ee
The Lie derivative along $\eta_{(2\ell)}$ of the background metric is found to have
the following non-trivial components
\bea
 \Lc_{\eta_{(2\ell)}}\gb_{tt}&=&\frac{2\ell(2\ell-1)\big(\G-k^2\g\big)r^2\rho_{(2\ell)}(t)
   -\big((2\ell-1)\G+k^2\g\big)\rho_{(2\ell)}''(t)}{\ell(2\ell-1)r^{2\ell-2}}\nn
 \Lc_{\eta_{(2\ell)}}\gb_{t\phi}=\Lc_{\eta_{(2\ell)}}\gb_{\phi t}&=&
   -\frac{k\g\Big(2\ell(2\ell-1)r^2\rho_{(2\ell)}(t)
      +\rho_{(2\ell)}''(t)\Big)}{2\ell(2\ell-1)r^{2\ell-1}}\nn
 \Lc_{\eta_{(2\ell)}}\gb_{rr}&=&\frac{(4\ell-4)\G\rho_{(2\ell)}(t)}{r^{2\ell}}
\eea
It follows that the corresponding charges $Q_{\eta_{(2\ell)}}$ all vanish.

Our final example concerns the fate of the original $U(1)$ isometry in conjunction
with the $SL(2,\mathbb{R})$-enhanced Virasoro algebra.
Since $\Lc_{\pa_\phi}\gb_{\mu\nu}=0$, it survives all consistent sets of boundary conditions.
We wish to demonstrate that it can be enhanced to a $U(1)$ current, though 
this current generates a spurious asymptotic symmetry.
To see this, let us weaken the boundary conditions (\ref{hrr}) and consider
\be
  h_{\mu\nu}=\left(\!\!\begin{array}{cccc} \Oc(r)&0&\Oc(1)&0 \\ 0&\Oc(r^{-3-2\ell_0})&0&0 \\ 
   \Oc(1)&0&0&0 \\ 0&0&0&0 \end{array} \!\!\right)
\label{hrr3}
\ee
preserving the centreless Virasoro algebra and the contravariant vector fields
$\eta_{(2\ell)}$ for $\ell\geq\ell_0+2$.
For every non-negative integer $\ell_0$, we have the vector field
\be
 \zeta=\psi(t)\pa_\phi
\ee
and for every $\ell\geq\ell_0$, we find the additional vector field
\be
 \zeta_{(2\ell)}=-\frac{\psi_{(2\ell)}'(t)}{(2\ell+\delta_{\ell,0})(2\ell+3)r^{2\ell+3}}\pa_t
   +\frac{\psi_{(2\ell)}(t)}{(2\ell+\delta_{\ell,0})r^{2\ell}}\pa_r
   +\frac{k\psi_{(2\ell)}'(t)}{(2\ell+\delta_{\ell,0})(2\ell+2)r^{2\ell+2}}\pa_\phi
\ee
We note that the exact $U(1)$ isometry follows from setting $\psi(t)=1$ in $\zeta$.
The non-trivial components of the corresponding Lie derivatives are
\be
 \Lc_{\zeta}\gb_{tt}=2k\g r\psi'(t),\qquad\quad 
 \Lc_{\zeta}\gb_{t\phi}=\Lc_{\zeta}\gb_{\phi t}=\g\psi'(t)
\label{Liezeta}
\ee
and
\bea
 \Lc_{\zeta_{(2\ell)}}\gb_{tt}&=&-\frac{(2\ell+2)(2\ell+3)\big(\G-k^2\g\big)r^2\psi_{(2\ell)}(t)
   -\big((2\ell+2)\G+k^2\g\big)\psi_{(2\ell)}''(t)}{(2\ell+\delta_{\ell,0})(\ell+1)(2\ell+3)r^{2\ell+1}}\nn
 \Lc_{\zeta_{(2\ell)}}\gb_{t\phi}= \Lc_{\zeta_{(2\ell)}}\gb_{\phi t}&=&
   \frac{k\g\Big((2\ell+2)(2\ell+3)r^2\psi_{(2\ell)}(t)+\psi_{(2\ell)}''(t)\Big)}{(2\ell+\delta_{\ell,0})
      (2\ell+2)(2\ell+3)r^{2\ell+2}}\nn
 \Lc_{\zeta_{(2\ell)}}\gb_{rr}&=&-\frac{2(2\ell+1)\G\psi_{(2\ell)}(t)}{(2\ell+\delta_{\ell,0})r^{2\ell+3}}
\eea
Now, weakening the boundary conditions (\ref{hrr3}) to (\ref{hw}), compatible with the Virasoro
generators $\xi$ (\ref{xi}), we find
\be
 \sqrt{-\gb}k_\zeta[h;\gb]\big|_{d\var\wedge d\theta}=-\sqrt{\frac{\al}{4\g\G}}k\g r\psi(t)\pa_\phi
   h_{r\phi}+\Oc(r^{-1})
\ee
where $h_{r\phi}=\Oc(r^{-1})$. However, we can ignore total $\phi$-derivatives, 
implying that $Q_{\zeta}=0$. One could, perhaps, still wonder if the formalism would allow a 
central extension when combining $\xi$ and $\zeta$. This is not the case, though, since we have
\be
 \sqrt{-\gb}k_\xi[\Lc_{\zeta}\gb;\gb]\big|_{d\phi\wedge d\theta}=
   \sqrt{\frac{\al\g^3}{\G}}\frac{\pa_r(r\xi^\phi)\psi'(t)}{4r}
 =\Oc(r^{-4})
\ee
while $\!\int\!\sqrt{-\gb}k_\zeta[\Lc_{\xi}\gb;\gb]=0$ since $Q_\zeta=0$ and $\xi$ 
is compatible with all the boundary conditions considered
here. We also find that $Q_{\zeta_{(2\ell)}}=0$.
\vskip.5cm
\subsection*{Acknowledgments}
\vskip.1cm
\noindent
This work is supported by the Australian Research Council. 
The author thanks Niels Obers (The Niels Bohr Institute)
for discussions sparking the undertaking of this work,
and The Niels Bohr Institute for its generous hospitality during his visit there
in May 2009. The author also thanks Dumitru Astefanesei for comments.


\end{document}